\documentclass{article}

\usepackage{amssymb,amsfonts,amsmath}
\usepackage{cite,enumerate,float}
\usepackage{color}
\usepackage{tikz}
\usetikzlibrary{arrows,snakes,backgrounds}
\usepackage{fancybox}
\usepackage{url}

\makeatletter
\DeclareFontEncoding{LS1}{}{}
\DeclareFontSubstitution{LS1}{stix}{m}{n}
\DeclareMathAlphabet{\mathcal}{LS1}{stixscr}{m}{n}
\makeatother

\def\be{\begin{eqnarray}}
\def\ee{\end{eqnarray}}
\def\nn{\nonumber}

\def\Tr{{\rm Tr}\,}

\def\MD{{\cal M}}%{M\!D}

\definecolor{red}{rgb}{1,0,0}
\definecolor{orange}{rgb}{1,0.5,0}
\definecolor{violet}{rgb}{0.7,0,1}

\hoffset= - 1.0in         % switch off for draft style
\begin{document}

\title{\vspace{1.5cm}\bf\Large
A note on universality in refined Chern-Simons theory
}

\author{
Andrei Mironov$^{a,b,c,}$\footnote{mironov@lpi.ru,~mironov@itep.ru},~~
Ruben Mkrtchyan$^{d,}$\footnote{mrl55@list.ru}
}

\date{ }

\maketitle

\vspace{-6cm}

\begin{center}
  \hfill FIAN/TD-07/26\\
  \hfill ITEP/TH-19/26\\
   \hfill IITP/TH-17/26
\end{center}

\vspace{4.5cm}

\begin{center}
 $^a$ {\small {\it Lebedev Physics Institute, Moscow 119991, Russia}}\\
$^b$ {\small {\it NRC "Kurchatov Institute", 123182, Moscow, Russia}}\\
$^c$ {\small {\it Institute for Information Transmission Problems, Moscow 127994, Russia}}\\
$^d${\small{\it Alikhanyan National Science Laboratory, Yerevan 0036, Armenia}}
\end{center}

\vspace{.1cm}

\begin{abstract}
We discuss various forms of refinements of Vogel’s universality in Chern–Simons theory. While the original universality applies to arbitrary simple Lie groups, its counterpart in refined Chern–Simons theory is restricted to simply laced Lie groups.
\end{abstract}

\bigskip

\newcommand\smallpar[1]{
  \noindent $\bullet$ \textbf{#1}
}

\paragraph{Introduction.}

Vogel's universality \cite{Vogel95,Vogel99} appears as a two-parameter generalization of simple Lie algebras into some Universal Lie Algebra, until the discovery \cite{Vogel11} of zero divisors in Vogel's $\Lambda$-algebra (see a modern review in \cite{KLS,KMS,MS}), then becoming the   unification tool for simple Lie algebras, particularly in their physics applications.

The main claim of universality is that there is a set of quantities basically associated  with knot invariants that can be realized universally (particularly for  all simple Lie algebras) at once as functions on the two-dimensional projective plane parameterized by three homogeneous (Vogel's, universal) parameters $\alpha$, $\beta$, $\gamma$. Since some knot invariants can be realized as Wilson loops in Chern-Simons theory with action $\frac{\kappa}{4\pi}\Tr\left({\cal A}d{\cal A} +
\frac{2}{3}{\cal A}^3\right)$  and hence are parameterized by the gauge group {\bf and} by the coupling constant $\kappa$, the universal quantities also depend on the parameter $q=\exp\left({2\pi i\over \kappa+h^\vee}\right)$, where $h^\vee$ is the dual Coxeter number. In fact, the universal quantities are functions of $q^\alpha$, $q^\beta$, $q^\gamma$, i.e. they are invariant w.r.t. simultaneous rescaling $(\alpha,\beta,\gamma)\to  (\xi\alpha,\xi\beta,\xi\gamma)$ and changing $q\to q^{\xi^{-1}}$, which for Lie algebras is equivalent to rescaling of invariant scalar product in algebra.
An important property due to P. Vogel \cite{Vogel95,Vogel99,Vogel11} is that the universal quantities are covariant w.r.t. the permutations of the Vogels' parameters.

The Vogel's parameters for simple Lie algebras (in the minimal normalization of the scalar product, i.e. when square of long root is 2, and $\alpha+\beta+\gamma=h^\vee$) are listed in Table \ref{vogelparm}.
\begin{table}[!ht]
\centering
\begin{tabular}{|c|c|c|c|c|c|}
\hline
Root system & Lie algebra & $\alpha$ & $\beta$ & $\gamma$ & $\alpha+\beta+\gamma$ \\
\hline
$A_n$ & ${sl}_{n+1}$ & $-2$ & $2$ & $n+1$ & $n+1$ \\
$B_n$ & ${so}_{2n+1}$ & $-2$ & $4$ & $2n-3$ & $2n-1$ \\
$C_n$ & ${sp}_{2n}$ & $-2$ & $1$ & $n+2$ & $n+1$ \\
$D_n$ & ${so}_{2n}$ & $-2$ & $4$ & $2n-4$ & $2n-2$ \\
$G_2$ & ${g}_2$ & $-2$ & $\frac{10}{3}$ & $\frac{8}{3}$ & $4$ \\
$F_4$ & ${f}_4$ & $-2$ & $5$ & $6$ & $9$ \\
$E_6$ & ${e}_6$ & $-2$ & $6$ & $8$ & $12$ \\
$E_7$ & ${e}_7$ & $-2$ & $8$ & $12$ & $18$ \\
$E_8$ & ${e}_8$ & $-2$ & $12$ & $20$ & $30$ \\
\hline
\end{tabular}
\caption{Vogel's parameters}
\label{vogelparm}
\end{table}

Among  universal quantities are: the Chern-Simons partition function \cite{MkrtVes12,Mkrt13,KreflMkrt,M2}, the dimensions \cite{Vogel99,LandMan06} and quantum dimensions \cite{Westbury03,MkrtQDims}, eigenvalues of the second and higher order Casimir operators \cite{LandMan06,MkrtSergVes,ManeIsaevKrivMkrt, IsaevProv, IsaevKriv,IsaevKrivProv}, the volume of compact simple Lie groups \cite{KhM}, trigonometric R-matrix \cite{W25}, the HOMFLY-PT knot/link polynomial colored with adjoint representation \cite{MMMuniv,W15,MMuniv,BMt4} and the Racah matrix involving the adjoint representation and its descendants \cite{MMuniv,ManeIsaevKrivMkrt, IsaevProv, IsaevKriv,IsaevKrivProv} (see also \cite{KLS}).

An extension of the universality to the refined Chern-Simons theory \cite{AgSh1,AgSh2,KS,AM1,R,AvMkrtString,Mane,AM2,AM3} introduces another parameter (coupling constant) $t$ in the case of simply laced groups, and a few parameters $t_a$'s otherwise. As was observed in \cite{KS,AM1,Mane}, the universality is preserved under refinement only in the simply laced case. A technical reason for this \cite{BM,B} is that, in the unrefined case, the universality is expressed in factorization properties of characters at the special point $q^{2\rho}$ ($\rho$ is the Weyl vector), and they are factorized also at the point $q^{2\rho^\vee}$ ($\rho^\vee$ is the dual Weyl vector), while, under the refinement, the factorization survives only at the point $q^{2r_k}$ and generally not at $q^{2\rho_k}$ ($\rho_k$ and $r_k$ are the refined versions of the Weyl vector and the dual Weyl vector correspondingly). Hence, the universality and factorization persists only in the simply laced case, when these two Weyl vectors coincide, $\rho_k=r_k$, and from now on we deal with the simply laced case only.

As we explain below, the universal quantities in the refined theory are no longer invariant w.r.t. permutations of the Vogels' parameters. However, one can naturally lift them to functions given on three-dimensional space of parameters that become symmetric functions of the three variables. Our main goal in this note is to clarify this point. We explain it, and then illustrate with examples worked out earlier in \cite{KS,AM1,Mane,BM,BMM,B}, collecting all the results about universality in refined Chern-Simons theory obtained so far at one place.

\bigskip

\paragraph{Notation.} Throughout the text, by refined dimension we always mean a refinement of the quantum dimension generated by a specialization of the Macdonald polynomial $P^R_\Lambda$, see details in \cite{BM,B}.

In variance with the original work by I. Macdonald \cite{Mac} and subsequent papers on the subject \cite{MacConj,CherednikConj,CherednikDAHA,Koorn},
we use symmetric quantum numbers, which allows us to present the results in a shorter and more elegant form: in our notation, the Macdonald polynomial depends on the squares of the original Macdonald parameters:
\begin{equation}
    q \,\rightarrow \, q^2, \,\, t \,\rightarrow \, t^2\nn
\end{equation}
and the symmetric brackets are defined to be
\be
\{x\}:=x-{1\over x},\ \ \ \ \ \ \ \ \ \ \{x\}_+:=x+{1\over x}\nn
\ee

\bigskip

\paragraph{General description.} In \cite{BM,BMM,B}, there were presented various quantities in refined Chern-Simons theory in a universal form suitable for the simply laced root systems. These quantities were evaluated basing on the Macdonald dimensions, i.e. values of the Macdonald polynomials $P^R$ associated with various root systems $R$ at specialization $t^{2\rho}$, where $\rho = (\rho_1, \dots,\rho_n)$ is the Weyl vector. By this we mean that one should make the substitution in the Macdonald polynomial depending on variables $x_1, \dots, x_n$:
\begin{equation}
    x_i = t^{2\rho_i}\nn
\end{equation}

For the simply laced root system $R$, which is our point of interest here, these Macdonald dimensions are factorized, and are given by formula \cite{Mac,CherednikConj}

\begin{equation}
P^R_\Lambda(t^{2\rho})=\prod_{\alpha\in R_+} \, \prod_{j=1}^{(\alpha,\Lambda)}\,  \frac{\left\{t^{(\rho,\alpha)+1}q^{j-1}\right\}}{\left\{t^{(\rho,\alpha)}q^{j-1}\right\}}\nn
\end{equation}
where $\Lambda$ is the dominant weight enumerating the Macdonald polynomial, and the notation of the roots $\alpha$ should not be confused with the Vogel's parameter.

Note that these Macdonald dimensions do not depend on normalization of the scalar product in the root system $(\bullet,\bullet)$.
Hence, all the quantities presented in the universal form effectively depend on the ratios of the Vogel's parameters $\beta/\alpha$ and $\gamma/\alpha$.

In fact, the results in \cite{BM,BMM,B} were presented as symmetric functions of variables
\be
u:=q^{\mathfrak{-2}},\ \ \ \ \ v:=t^{\mathfrak{-2\beta/\alpha}},\ \ \ \ \ w:=t^{\mathfrak{-2\gamma/\alpha}}\nn
\ee

and also variables $q,t$. It is also convenient to introduce   $T:={q^2\over t^2}uvw$

We would like to consider these functions at an arbitrary independent values of $u,v,w, q, t$. This means that each universal refined quantity is associated with a function of three variables $u$, $v$, $w$ and of two parameters $q$, $t$: $F(u,v,w;q,t)$. We call this function {\bf generalized universal quantity}. It enjoys the properties

\begin{itemize}
\item $F(u,v,w;q,t)$ is invariant (covariant) under permutations of $u$, $v$, $w$.

\item $F(q^\alpha,q^\beta,q^\gamma;q,q)$ is equal to the corresponding unrefined quantity (which is a function of $q^{\alpha}$, $q^{\beta}$, $q^{\gamma}$ only).

\item The reduction $F(q^{-2},t^{-2\beta/\alpha},t^{-2\gamma/\alpha};q,t)$ gives rise to a universal function, equal to the initial refined quantity.  It is clearly invariant w.r.t. the rescaling of universal parameters, but is invariant (covariant) under permutation $\beta \leftrightarrow \gamma $, only. One can bring this function into a more standard form by redefinition of $q$ and $t$, $q\to q^{-\alpha/2}$, $t\to t^{-\alpha/2}$, then it becomes $F(q^{\alpha},t^{\beta},t^{\gamma};q^{-\alpha/2},t^{-\alpha/2})$,
    which clearly is invariant under  rescaling of all universal  parameters by a factor of $\xi$ and simultaneous change $q\to q^{\xi^{-1}}$ and $t\to t^{\xi^{-1}}$. Note that with these redefinitions $T=t^{\alpha+\beta+\gamma}$.

     We call this function {\bf Vogel's universal refined quantity}. At $q=t$ it is equal to the initial unrefined quantity.

\end{itemize}

Now we consider all examples of refined universal expressions known so far. From now on, we use gothic letters for the generalized universal quantities, e.g.  $\mathfrak{F}, \widetilde{\mathfrak{F}}$, and the calligraphic ones, e.g.  $\cal{F}, \widetilde{\cal{F}}$,   for the Vogels' universal quantities in order to distinguish between them. Thus, the generalized universal function $\mathfrak{F}(u,v,w;q,t)$ gives rise to the Vogel's universal function

\be
\boxed{
{\cal F}(\alpha,\beta,\gamma;q,t)=\mathfrak{F}(q^{\alpha},t^{\beta},t^{\gamma};q^{-\alpha/2},t^{-\alpha/2})
}\nn
\ee

\bigskip

\paragraph{Universal refined adjoint dimension $\mathfrak{M}_{Adj}$.}

The simplest example  is the refined adjoint dimension. The generalized universal refined adjoint dimension is \cite{BM,B}
\be\label{adj}
    \mathfrak{M}_{Adj}=-{\left\{\frac{T}{\sqrt{u}}\right\}\left\{\frac{T}{\sqrt{v}}\right\}\left\{\frac{T}{\sqrt{w}}\right\}\over
    \{\sqrt{u}\}
    \{\sqrt{v}\}\{\sqrt{w}\}}\cdot{\xi(T)\over \xi(t)}
\ee
where
\be
\xi(x):={\{x\}\over\{qx/t\}}\stackrel{t=q}{\longrightarrow}1\nn
\ee
This expression is invariant w.r.t. permutations of $u$, $v$, and $w$.

At the same time, the Vogel's universal refined dimension, in accordance with the general rule, is

\be
{\cal M}_{Adj}=-{\{t^{\alpha+\beta+\gamma}\}\over\{q^{-{1\over 2}\alpha}t^{{3\over 2}\alpha+\beta+\gamma}\}}
{\{q^{-{1\over 2}\alpha}t^{\alpha+\beta+\gamma}\}\{t^{\alpha+{1\over 2}\beta+\gamma}\}\{t^{\alpha+\beta+{1\over 2}\gamma}\}
\over\{t^{{1\over 2}\alpha}\}\{t^{{1\over 2}\beta}\}\{t^{{1\over 2}\gamma}\}}\nn
\ee
and this expression is invariant only w.r.t. permutation of $\beta$ and $\gamma$. 

\bigskip

\paragraph{Universal decomposition of $\left(\mathfrak{M}_{Adj}\right)^2$.}

The universal decomposition of the square of the universal adjoint dimension in the refined case \cite{BMM}, similarly to the unrefined one \cite{Vogel95,Vogel99}, is a sum of six universal refined dimensions associated with six ``universally-irreducible" entities
({\bf uirreps} representations, in accordance with \cite{BMM}). However, in variance with the unrefined case, this decomposition is not given just by a sum of refined dimensions of separate representations: the square of Macdonald polynomials expands with non-trivial Littlewood-Richardson coefficients that are rational functions of $q$ and $t$. As was explained in \cite{BMM}, universal are, in this case, only the combinations of the Littlewood-Richardson coefficients along with the Macdonald dimensions. Nevertheless, these objects are enough not only for obtaining this decomposition, but also for constructing universal hyperpolynomials of the $T[2,2n]$ links. Since universality is expected to have something to do only with Chern-Simons theory observables, i.e. with knot invariants, these combined objects provide all necessary building blocks for universal quantities.

Speaking more concretely, the generalized universal decomposition formula looks like (for the details, see \cite{BMM})
\be\label{du1}
\left(\mathfrak{M}_{Adj}\right)^2=\mathfrak{X}_2+\mathfrak{Y}_2(u)+\mathfrak{Y}_2(v)+\mathfrak{Y}_2(w)
+\mathfrak{P}_{Adj}+\mathfrak{P}_\emptyset
\ee
where the separate terms of the decomposition are

\bigskip

\be\label{du2}
    \mathfrak{X}_2&=&-\mathfrak{M}_{Adj}\times{\left\{\frac{q}{t}\sqrt{Tu}\right\}\left\{\frac{q}{t}\sqrt{Tv}\right\}
    \left\{\frac{q}{t}\sqrt{Tw}\right\}\left\{\frac{T}{u}\right\}\left\{\frac{T}{v}\right\}\left\{\frac{T}{w}\right\}\over
    \left\{\frac{qu}{t}\right\}\left\{\frac{qv}{t}\right\}\left\{\frac{qw}{t}\right\}
    \left\{\sqrt{\frac{T}{u}}\,\right\}\left\{\sqrt{\frac{T}{v}}\,\right\}\left\{\sqrt{\frac{T}{w}}\,\right\}}
   \cdot
      {\xi(qT)\xi\left(t\sqrt{T}\right)\xi\left(\frac{\sqrt{T}}{t}\right)\over
    \xi(t^2T)\xi(qtT)}\cdot{\xi(t^2)\over\xi(t)\xi\left(\frac{1}{t^2}\right)}
    \nn\\ \nn \\
\mathfrak{Y}_2(u)&=&
{\left\{\frac{T}{u\sqrt{u}}\right\}\left\{\frac{T}{\sqrt{uv}}\right\}\left\{\frac{T}{\sqrt{uw}}\right\}
\left\{\frac{T}{\sqrt{v}}\right\}\left\{\frac{T}{\sqrt{w}}\right\}\{T\}
\over
\left\{ u \right\} \left\{\sqrt{ u} \right\}\left\{\sqrt{ v} \right\}\left\{\sqrt{ w} \right\}\left\{\sqrt{\frac{v}{ u}} \right\}\left\{\sqrt{ \frac{w}{u}} \right\}}
\cdot \frac{\xi\left(\frac{T}{{u}}\right) \xi\left(\frac{T}{\sqrt{u}}\right)\xi\left(u\right)  }{\xi\left(t\right)^2\xi\left(\sqrt{u}\right)}
\nn\\
\mathfrak{Y}_2(v)&=&\mathfrak{Y}_2(u)\Big|_{u\leftrightarrow v}
\nn\\
\mathfrak{Y}_2(w)&=&\mathfrak{Y}_2(u)\Big|_{u\leftrightarrow w}
\nn\\
\mathfrak{P}_\emptyset&=&{\xi\left(\frac{1}{q}\right)\over\xi(t)}\cdot {\xi(T)
\xi\left(\frac{T}{\sqrt{u}}\right) \xi\left(\frac{T}{\sqrt{v}}\right)\xi\left(\frac{T}{\sqrt{w}}\right)\over
\xi(qT/t)
\xi(\sqrt{u})\xi(\sqrt{v})\xi(\sqrt{w})}
\ee
The remaining term $\mathfrak{P}_{Adj}=C_{Adj}\cdot \MD_{Adj}$, $\MD_{Adj}$ being ($u$,$v$,$w$)-symmetric, and such being also the coefficient $C_{Adj}$. However, $C_{Adj}$ is quite involved, and one
can restore $\mathfrak{P}_{Adj}$ in the universal form in the simplest way from formulas (\ref{du1})-(\ref{du2}). One can see that, indeed, (\ref{du1}) is a covariant decomposition w.r.t. permutations of $u$, $v$, and $w$.

\bigskip

At the same time, the Vogel's universal decomposition in this case is
\be\label{duV1}
\left(\MD_{Adj}\right)^2={\cal X}_2+{\cal Y}^\alpha_2+{\cal Y}^\beta_2+{\cal Y}^\gamma_2
+{\cal P}_{Adj}+{\cal P}_\emptyset
\ee
and ${\cal Y}^{\beta,\gamma}_2$ are not obtained by a simple permutations from ${\cal Y}^\alpha_2$. The separate terms of the decomposition are

\bigskip

\be\label{duV2}
    {\cal X}_2&=&-{\cal M}_{Adj}\times{\{t^{\alpha+{\beta\over 2}+{\gamma\over 2}}\}
    \{q^{-{\alpha\over 2}}t^{\alpha+{\beta\over 2}+\gamma}\}\{q^{-{\alpha\over 2}}t^{\alpha+\beta+{\gamma\over 2}}\}
    \over
    \{q^{\alpha\over 2}t^{\alpha\over 2}\}\{q^{-{\alpha\over 2}}t^{{\alpha\over 2}+\beta}\}\{q^{-{\alpha\over 2}}t^{{\alpha\over 2}+\gamma}\}}
    \{q^{-{\alpha\over 2}}t^{{\alpha\over 2}+{\beta\over 2}+{\gamma\over 2}}\}_+
    \{t^{{\alpha\over 2}+{\beta\over 2}}\}_+\{t^{{\alpha\over 2}+{\gamma\over 2}}\}_+\times\nn\\
    &\times&
    {\zeta(t^{\alpha+{\beta\over 2}+{\gamma\over 2}})\zeta(q^{-{\alpha\over 2}}t^{\alpha+\beta+\gamma})\zeta(q^{\alpha\over 2})\zeta(t^{{\beta\over 2}+{\gamma\over 2}})\over\zeta(t^{\beta+\gamma})\zeta(q^{-{\alpha\over 2}}t^{{\alpha\over 2}+\beta+\gamma})\zeta(t^\alpha)\zeta(q^{\alpha\over 2}t^{\alpha\over 2})}=
    \nn\\
   &=&-{\cal M}_{Adj}\times{\{t^{\alpha+{1\over 2}\beta+{1\over 2}\gamma}\}^2
    \{q^{-{1\over 2}\alpha}t^{\alpha+{1\over 2}\beta+\gamma}\}\{q^{-{1\over 2}\alpha}t^{\alpha+\beta+{1\over 2}\gamma}\}
    \{q^{-{1\over 2}\alpha}t^{\alpha+\beta+\gamma}\}\over
    \{q^{-\alpha}t^{{3\over 2}\alpha+\beta+\gamma}\}\{q^{-{1\over 2}\alpha}t^{{3\over 2}\alpha+{1\over 2}\beta+{1\over 2}\gamma}\}
    \{q^{-{1\over 2}\alpha}t^{{1\over 2}\alpha+\beta}\}\{q^{-{1\over 2}\alpha}t^{{1\over 2}\alpha+\gamma}\}}\times\nn\\
    &\times&{\{q^{{1\over 2}\alpha}\}\{q^{-{1\over 2}\alpha}t^{{3\over 2}\alpha}\}\over
    \{q^{{1\over 2}\alpha}t^{{1\over 2}\alpha}\}^2\{t^{{1\over 2}\alpha}\}}
    {\{q^{-{1\over 2}\alpha}t^{{1\over 2}\alpha+{1\over 2}\beta+{1\over 2}\gamma}\}^2_+
    \{t^{{1\over 2}\alpha+{1\over 2}\beta}\}_+\{t^{{1\over 2}\alpha+{1\over 2}\gamma}\}_+\over \{t^{{1\over 2}\beta+{1\over 2}\gamma}\}_+}
    \nn\\
    \nn \\
{\cal Y}^\alpha_2&=&
{\{q^{-{3\over 2}\alpha}t^{\alpha+\beta+\gamma}\}\{q^{-{\alpha\over 2}}t^{\alpha+{\beta\over 2}+\gamma}\}
\{q^{-{\alpha\over 2}}t^{\alpha+\beta+{\gamma\over 2}}\}\{t^{\alpha+{\beta\over 2}+\gamma}\}
\{t^{\alpha+\beta+{\gamma\over 2}}\}\{t^{\alpha+\beta+\gamma}\}
\over\{q^{{\alpha\over 2}}t^{\alpha\over 2}\}\{t^{\alpha\over 2}\}\{t^{\beta\over 2}\}\{t^{\gamma\over 2}\}
\{q^{-{\alpha\over 2}}t^{\beta\over 2}\}\{q^{-{\alpha\over 2}}t^{\gamma\over 2}\}}\times\nn\\
&\times& \zeta(q^{-\alpha}t^{\alpha+\beta+\gamma}\}\zeta(q^{-{\alpha\over 2}}t^{\alpha+\beta+\gamma})
\nn\\ \nn\\
{\cal Y}^\beta_2&=&{\{t^{\alpha-{\beta\over 2}+\gamma}\}\{q^{-{\alpha\over 2}}t^{\alpha+{\beta\over 2}+\gamma}\}
\{t^{\alpha+{\beta\over 2}+{1\over 2}\gamma}\}t^{\alpha+{\beta\over 2}+\gamma}\}\{t^{\alpha+\beta+\gamma}\}
\{q^{-{\alpha\over 2}}t^{\alpha+\beta+\gamma}\}\over\{q^{\alpha\over 2}\}\{t^{\beta\over 2}\}\{t^{\gamma\over 2}\}
\{q^{\alpha\over 2}t^{-{\beta\over 2}}\}\{t^{{\gamma\over 2}-{\beta\over 2}}\}\{q^{-{\alpha\over 2}}t^{{\alpha\over 2}+\beta}\}}
\times{\zeta(t^{\alpha+\gamma})\zeta(t^{\alpha+{1\over 2}\beta+\gamma})\over\zeta(t^{\beta\over 2})\zeta(t^{-{\alpha\over 2}})^2}
\nn\\ \nn\\
{\cal Y}^\gamma_2&=&{\cal Y}^\beta_2\Big|_{\beta\leftrightarrow \gamma}
\nn\\ \nn\\
{\cal P}_\emptyset&=&{\zeta(t^{\alpha+\beta+\gamma})\zeta(q^{-{\alpha\over 2}}t^{\alpha+\beta+\gamma})
\zeta(t^{\alpha+{1\over 2}\beta+\gamma})\zeta(t^{\alpha+\beta+{1\over 2}\gamma})\over
\zeta(q^{-{\alpha\over 2}}t^{{3\over 2}\alpha+\beta+\gamma})
\zeta(t^{-{\alpha\over 2}})\zeta(t^{\beta\over 2})\zeta(t^{\gamma\over 2})}
\ee
where
\be
\zeta(x):={\{x\}\over\{q^{-{\alpha\over 2}}t^{\alpha\over 2}x\}}\stackrel{t=q}{\longrightarrow}1\nn
\ee
and (\ref{duV1}) is a covariant decomposition only w.r.t. permutations of $\beta$ and $\gamma$.

\bigskip

\paragraph{Universal Taki factor.}
The framing factor of knot/link invariants at representation given by the highest weight vector $\Lambda$ is given by $q^{nC_2(\Lambda)}$, where $n$ is an integer and $C_2(\Lambda)$ is the eigenvalue of the second Casimir operator,
\be
C_2(\Lambda)=(\Lambda,\Lambda)/2+(\Lambda,\rho)\nn
\ee
Universal formulas for this quantity in the adjoint representation as well as in representations emerging in decompositions of its powers are available in the unrefined case \cite{Vogel95,Vogel99}.

The refinement of $q^{C_2(\Lambda)}$ is called Taki factor \cite{Taki}, and can be written for any simply laced root system in the form
\be
f_\Lambda=q^{(\Lambda,\Lambda)/2}t^{(\Lambda,\rho)}\nn
\ee
This factor also admits the generalized universal expressions:

\be\label{taki1}
   & \mathfrak{f}_{Adj} = \frac{q}{t}T\nn\\
   & \mathfrak{f}_{X_2}  = \frac{q^3}{t^3} T^2\nn  \\
   & \mathfrak{f}_{Y_2(u)} = {q^2\over t^2}{T^2\over u} \\
   & \mathfrak{f}_{Y_2(v)} = {q^2\over t^2}{T^2\over v}\nn \\
   & \mathfrak{f}_{Y_2(w)} =  {q^2\over t^2}{T^2\over w} \nn\\
   & \mathfrak{f}_{\varnothing} = 1\nn
\ee

Similarly, the Vogels' universal expressions are

\be\label{taki2}
   & {\cal f}_{Adj} = q^{-{1\over 2}\alpha}t^{\beta+\gamma+{3\over 2}\alpha}\nn\\
   & {\cal f}_{X_2}  = q^{-{3\over 2}\alpha}t^{2\beta+2\gamma+{7\over 2}\alpha}\nn  \\
   & {\cal f}_{Y_2^\alpha} = q^{-2\alpha}t^{2\beta+2\gamma+3\alpha} \\
   & {\cal f}_{Y_2^\beta} = q^{-\alpha}t^{\beta+2\gamma+3\alpha}\nn \\
   & {\cal f}_{Y_2^\gamma} = q^{-\alpha}t^{2\beta+\gamma+3\alpha} \nn\\
   & {\cal f}_{\varnothing} = 1\nn
\ee

\bigskip

\paragraph{Universal Hopf hyperpolynomial.}
Now we have all necessary invariant quantities at hands for constructing the link invariant, the hyperpolynomial of $T[2,2n]$ link \cite{BMM}. It has the generalized universal form

\bigskip

\be\label{linku}
\mathfrak{H}_{Adj,Adj}^{T[2,2n]} =\mathfrak{f}_{Adj}^{4n}\left[\mathfrak{f}_{X_2}^{-2n}\mathfrak{X}_2
+\mathfrak{f}_{Y_2(\mathfrak{a})}^{-2n}\mathfrak{Y}_2(\mathfrak{a})+
\mathfrak{f}_{Y_2(\mathfrak{b})}^{-2n}\mathfrak{Y}_2(\mathfrak{b})+\mathfrak{f}_{Y_2(\mathfrak{c})}^{-2n}\mathfrak{Y}_2(\mathfrak{c})
+\mathfrak{f}_{Adj}^{-2n}\mathfrak{P}_{Adj}+\mathfrak{P}_\emptyset\right]
\ee
with all necessary quantities evaluated in (\ref{du2}), (\ref{taki1}), and similarly in the Vogel's universal form, in (\ref{duV2}), (\ref{taki2}). At $n=1$, formula (\ref{linku}) reduces to the Hopf link hyperpolynomial, which, for any root system, admits an explicit realization both in terms of Macdonald polynomials evaluated at special points \cite[Eq.(51)]{BMM} and in terms of integrals, these two are related \cite[Eq.(56)]{BMM} by the CMM (Cherednik-Macdonald-Mehta) formulas \cite{Che,EK,ChaE,MMP}. As was checked in \cite{BMM}, both these types of formulas match the universal formula (\ref{linku}), however, their formulation in the universal terms is not available.

\bigskip

\paragraph{Universal partition function.} In accordance with \cite{KS,AM1,Mane}, the refined partition function of Chern-Simons on the $S^3$ in the simply laced case can be rewritten as the generalized universal quantity\footnote{At $t=q^k$ with an integer $k$, this partition function for the root system $R$ is proportional \cite{AgSh1,AgSh2} to (a very particular case of) the CMM formula:
\begin{equation}
\oint \prod_i{dx_i\over x_i}\Delta^R_+(x)\Delta^R_-(x)\sum_{\Lambda \in P} q^{(\Lambda,\Lambda)+2(\Lambda,z)}
= |W_R|\ \Delta_{+}^R (q^{2\rho_k})\nn
\end{equation}
where $|W_R|$ is order of the Weyl group $W_R$ of the root system $R$, the sum goes over the whole weight lattice $P = \mathbb{Z}\,\omega$, and
\begin{equation}
    \Delta^R_{\pm}(x):\ \stackrel{x_i = q^{2z_i}}{=}\  \prod_{\alpha \in R_{\pm}} \prod_{j=0}^{k-1} \{q^{(\alpha,z)+j}\}\nn
\end{equation}
where the product runs over all positive or negative roots accordingly.
}

\be
\mathfrak{Z}(u,v,w;q,t)= \left(\log T\right)^{(\widetilde{\mathfrak{M}}_{Adj})_0}\exp\left(-\int_0^\infty{dx\over x(e^x-1)}\Big[\widetilde{\mathfrak{M}}_{Adj}(u^x,v^x,w^x,q^x,t^x)-\widetilde{\mathfrak{M}}_{Adj}(u^{\xi x},v^{\xi x},w^{\xi x},q^{\xi x},t^{\xi x})\Big]\right)\nn
\ee
where the function
\be 
   \widetilde{\mathfrak{M}}_{Adj}(u,v,w,q,t)=
   -{\left\{\frac{T}{\sqrt{u}}\right\}\left\{\frac{T}{\sqrt{v}}\right\}\left\{\frac{T}{\sqrt{w}}\right\}\over
    \{\sqrt{u}\}
    \{\sqrt{v}\}\{\sqrt{w}\}}\nn
\ee
has to be compared with (\ref{adj}): it differs from $\mathfrak{M}_{Adj}$ by a simple factor of ${\xi(T)\over \xi(t)}$.

$(\widetilde{\mathfrak{M}}_{Adj})_0$ is defined as
\be
(\widetilde{\mathfrak{M}}_{Adj})_0:=        \lim_{x\to 0}     \widetilde{\mathfrak{M}}_{Adj}(u^x,v^x,w^x,q^x,t^x)\nn
\ee

Choosing the regularizing parameter $\xi$ fixes the normalization of the partition function. Following \cite{KS,AM1,Mane}, we choose it to be
\be\label{xi}
\xi= {1\over \log T}
\ee
and $\mathfrak{Z}(u,v,w;q,t)=1$ as $T=1$.

\bigskip

Above quantities can be rewritten in the Vogel's universal form.  It is easy to show  that

\be
 \widetilde{\cal{M}}_{Adj}(\alpha,\beta,\gamma,q,t)  \equiv   \widetilde{\mathfrak{M}}_{Adj}(q^\alpha,t^\beta,t^\gamma;q^{-\alpha/2},t^{-\alpha/2})=
 -{\{q^{-{1\over 2}\alpha}t^{\alpha+\beta+\gamma}\}\{t^{\alpha+{1\over 2}\beta+\gamma}\}\{t^{\alpha+\beta+{1\over 2}\gamma}\}
\over\{t^{{1\over 2}\alpha}\}\{t^{{1\over 2}\beta}\}\{t^{{1\over 2}\gamma}\}}\nn
\ee
and partition function becomes:
\be
{\cal Z}(\alpha,\beta,\gamma;q,t)=\left(\log T\right)^{(\widetilde{\cal{M}}_{Adj})_0 }\exp\left(-\int_0^\infty{dx\over x(e^x-1)}\Big[\widetilde{\cal {M}}_{Adj}(x\alpha,x\beta,x\gamma;q,t)-\widetilde{\cal {M}}_{Adj}(\xi x\alpha,\xi x\beta,\xi x\gamma;q,t)\Big]\right)\nn
\ee
reproducing the original result of \cite{KS,AM1,Mane}. 

Note that, in this case, parameterizing in accordance with \cite{AgSh1,AgSh2} $q=\exp\left({2\pi i\over \kappa+yh^\vee}\right)$, $t=\exp\left({2\pi iy\over \kappa+yh^\vee}\right)$ ($y$ is the refinement parameter), one obtains  for $\xi$ in (\ref{xi}) \cite{KS,AM1,Mane}
\be
\xi= 1+\kappa{\log q\over \log T}\nn
\ee
i.e. the partition function is normalized to be unity at vanishing coupling constant $\kappa$ \cite{Wit}.

\bigskip

\paragraph{Concluding remarks.}

In this note, in the case of refined Chern-Simons theory with simply laced gauge groups, we presented examples of the generalized universal quantities covariant with respect to permutations of three parameters $u=q^\alpha$, $v=t^\beta$, $w=t^\gamma$, and of the Vogel's universal quantities, which are covariant only with respect to permutation of parameters $\beta$ and $\gamma$, which is natural, since restricting to simply laced algebras introduces additional structure that distinguishes $\alpha$ from the other parameters. This open the possibility of additional universal expressions. For instance \cite{AM1}, for simply laced algebras one has for the dimension of the adjoint representation the formulae
\begin{eqnarray} \nonumber
	dim_{Adj}= -	\frac{(\alpha+2\beta+2\gamma)(2\alpha+\beta+2\gamma)(2\alpha+2\beta+\gamma)}{\alpha\beta\gamma}=	\\ \label{rank}
		\frac{(\alpha-2(\alpha+\beta+\gamma))}{\alpha}	\frac{(2\alpha+\beta+2\gamma)(2\alpha+2\beta+\gamma)}{\beta\gamma} =
	(1+h^\vee)r
\end{eqnarray}
where $r$ is the rank of the algebra. Since $(2\alpha+2\beta+2\gamma)/(-\alpha)=h^\vee$,  (\ref{rank}) gives the  universal expression for the rank of the algebra.  Thus, when restricting to simply laced algebras, the rank becomes a universal quantity.

\bigskip

\paragraph{Acknowledgments.}

We are grateful to Maneh Avetisyan, Liudmila Bishler and Alexei Morozov for discussions and collaboration. The work of A.M. was partially funded within the state assignment of the Institute for Information Transmission Problems of RAS, and partly supported by the grant of the Foundation for the Advancement of Theoretical Physics and Mathematics “BASIS”, and by the grant 24WS-1C031 of  the Science Committee of the Ministry of Science and Education of the Republic of Armenia.  The work of R.M. is  partially supported by the Science Committee of the Ministry of Science and Education of the Republic of Armenia under contracts 21AG-1C060 and 24WS-1C031.


\bigskip

\begin{thebibliography}{12}

\bibitem{Vogel95} P. Vogel, {\sl Algebraic structures on modules of diagrams}, Preprint (1995), available at \url{https://webusers.imj-prg.fr/~pierre.vogel/diagrams.pdf},

\bibitem{Vogel99}  P. Vogel, {\sl The Universal Lie algebra}, Preprint (1999), available at \url{https://webusers.imj-prg.fr/~pierre.vogel/grenoble-99b.pdf}

\bibitem{Vogel11}  P. Vogel, {\sl Algebraic structures on modules of diagrams}, J. Pure Appl. Algebra, {\bf 215} (2011) 1292-1339



\bibitem{KLS} D.~Khudoteplov, E.~Lanina, A.~Sleptsov,
{\sl Construction of Lie algebra weight system kernel via Vogel algebra},
arXiv:2411.14417

\bibitem{KMS} D.~Khudoteplov, A.~Morozov, A.~Sleptsov,
{\sl Can Yang-Baxter imply Lie algebra?},
arXiv:2503.13437

\bibitem{MS} A.~Morozov and A.~Sleptsov,
%``Vogel{\textquoteright}s universality and the classification problem for Jacobi identities,''
Eur. Phys. J. \textbf{C85} (2025) 1233,
%doi:10.1140/epjc/s10052-025-14943-y
arXiv:2506.15280

\bibitem{MkrtVes12} R.L. Mkrtchyan, A.P. Veselov, %Universality in Chern-Simons theory.
JHEP {\bf 08} (2012) 153, arXiv:1203.0766

\bibitem{Mkrt13} R.L. Mkrtchyan, %Nonperturbative universal Chern-Simons theory,
JHEP {\bf 09} (2013) 54, arXiv:1302.1507

\bibitem{KreflMkrt} D. Krefl, R. Mkrtchyan, %Exact Chern-Simons / Topological String duality,
JHEP {\bf 10} (2015) 45, arXiv:1506.03907

\bibitem{M2} R.L.~Mkrtchyan,
%``Partition function of Chern\textendash{}Simons theory as renormalized $q$-dimension,''
J. Geom. Phys. \textbf{129} (2018) 186-191,
%doi:10.1016/j.geomphys.2018.03.009
arXiv:1709.03261

\bibitem{LandMan06} J.M. Landsberg, L. Manivel, %A universal dimension formula for complex simple Lie algebras.
Adv. Math. {\bf 201} (2006) 379-407

\bibitem{Westbury03} B. Westbury, %Invariant tensors and diagrams,
Proceedings of the Tenth Oporto Meeting on Geometry, Topology and Physics, {\bf 18} (2001) suppl. 2003, pp. 49-82

\bibitem{MkrtQDims} R.L. Mkrtchyan, %On universal quantum dimensions,
Nucl. Phys. {\bf B921} (2017) 236-249, arXiv:1610.09910



\bibitem{MkrtSergVes} R.L. Mkrtchyan, A.N. Sergeev, A.P. Veselov, %Casimir eigenvalues for universal Lie algebra,
J. Math. Phys. {\bf 53} (2012) 102106, arXiv:1105.0115

\bibitem{ManeIsaevKrivMkrt} M.Y. Avetisyan, A.P. Isaev, S.O. Krivonos, R.L. Mkrtchyan, %The uniform structure of $\mathfrak{g}^{\otimes4}$,
Russian Journal of Mathematical Physics, {\bf 31} (2024) 379-388, arXiv:2311.05358

\bibitem{IsaevProv} A.P. Isaev, A.A. Provorov, %Projectors on invariant subspaces of representations ad$^2$ of Lie algebras so(N) and sp(2r) and Vogel parametrization,
Theor.Math.Phys. {\bf 206} (2021) 3-22, arXiv:2012.00746

\bibitem{IsaevKriv} A.P. Isaev, S.O. Krivonos, %Split Casimir operator for simple Lie algebras, solutions of Yang-Baxter equations and Vogel parameters,
 J. Math. Phys. {\bf 62} (2021) 083503, arXiv:2102.08258

\bibitem{IsaevKrivProv} A.P. Isaev, S.O. Krivonos, A.A. Provorov, %Split Casimir operator for simple Lie algebras in the cube of ad-representation and Vogel parameters,
    Int. J. Mod. Phys. {\bf A38} (2023) 235003, arXiv:2212.14761

\bibitem{KhM} H.M.~Khudaverdian, R.L.~Mkrtchyan,
%``Universal volume of groups and anomaly of Vogel\textquoteright{}s symmetry,''
Lett. Math. Phys. \textbf{107} (2017) 1491-1514,
%doi:10.1007/s11005-017-0949-8
arXiv:1602.00337

\bibitem{W25} Bruce W. Westbury, Paul Zinn-Justin, A uniform trigonometric R-matrix for the exceptional series, arXiv:2406.01348,
https://doi.org/10.48550/arXiv.2406.01348

\bibitem{MMMuniv} A. Mironov, R. Mkrtchyan, A. Morozov, %On universal knot polynomials,
JHEP {\bf 02} (2016) 078, arXiv:1510.05884

\bibitem{W15}
B.W.Westbury, Extending and quantising the Vogel plane, arXiv:1510.08307.


\bibitem{MMuniv} A. Mironov, A. Morozov, %Universal Racah matrices and adjoint knot polynomials. I. Arborescent knots,
Phys. Lett. {\bf B755} (2016) 47-57, arXiv:1511.09077

\bibitem{BMt4} L.~Bishler and A.~Mironov,
%``Torus knots in adjoint representation and Vogel{\textquoteright}s universality,''
Eur. Phys. J. \textbf{C85} (2025) 911,
%doi:10.1140/epjc/s10052-025-14651-7
arXiv:2506.06219

\bibitem{AgSh1} M.~Aganagic, S.~Shakirov,
%``Knot Homology and Refined Chern-Simons Index,''
Commun. Math. Phys. \textbf{333} (2015) 187-228,
%doi:10.1007/s00220-014-2197-4
arXiv:1105.5117

\bibitem{AgSh2} M.~Aganagic, S.~Shakirov,
%``Refined Chern-Simons Theory and Knot Homology,''
Proc. Symp. Pure Math. \textbf{85} (2012) 3-32,
%doi:10.1090/pspum/085/1372
arXiv:1202.2489

\bibitem{KS} D. Krefl, A. Schwarz, %Refined Chern-Simons versus Vogel universality,
Journal of Geometry and Physics, {\bf 74} (2013) 119-129, arXiv:1304.7873

\bibitem{AM1} M.Y.~Avetisyan, R.L.~Mkrtchyan,
%``On partition functions of refined Chern-Simons theories on S$^{3}$,''
JHEP \textbf{10} (2021) 033,
%doi:10.1007/JHEP10(2021)033
arXiv:2107.08679

\bibitem{R} R.L.~Mkrtchyan,
%``Chern-Simons theory with the exceptional gauge group as a refined topological string,''
Phys. Lett. \textbf{b808} (2020) 135692,
%doi:10.1016/j.physletb.2020.135692
arXiv:2007.09346

\bibitem{AvMkrtString} M.Y. Avetisyan, R.L. Mkrtchyan, % On refined Chern-Simons/topological string duality for classical gauge groups.
JHEP {\bf 2022} (2022) 97, arXiv:2205.12832

\bibitem{Mane} M. Avetisyan, {\sl Vogel's Universality and its Applications}, arXiv:2207.04302

\bibitem{AM2} M.Y.~Avetisyan, R.L.~Mkrtchyan,
%``Two-fold refinement of non simply laced Chern-Simons theories,''
J. Geom. Phys. \textbf{191} (2023) 104907,
%doi:10.1016/j.geomphys.2023.104907
arXiv:2302.14319

\bibitem{AM3} M.Y.~Avetisyan, R.L.~Mkrtchyan,
%``Refined $E_n$ Chern-Simons theory,''
Phys. Part. Nucl. \textbf{54} (2023) 1059-1062,
arXiv:2304.05184

\bibitem{BM} L. Bishler, A. Mironov, %``On Refined Vogel's universality,''
Phys.Lett. {\bf B867} (2025) 139596, arXiv:2504.13831

\bibitem{B} L.~Bishler,
%``Vogel's universality and Macdonald dimensions,''
Nucl. Phys. \textbf{B1018} (2025) 117085,
%doi:10.1016/j.nuclphysb.2025.117085
arXiv:2507.11414

\bibitem{BMM} L.~Bishler, A.~Mironov and A.~Morozov,
%``Macdonald deformation of Vogel's universality and link hyperpolynomials,''
Phys. Lett. \textbf{B868} (2025) 139695,
%doi:10.1016/j.physletb.2025.139695
arXiv:2505.16569

\bibitem{Mac} I.G. Macdonald, %Orthogonal polynomials associated with root systems,
S\`eminaire Lotharingien Combin. {\bf 45} (2000), Article B45a, 40 pp, arXiv:math/0011046

\bibitem{CherednikConj} I. Cherednik, %Macdonald's Evaluation Conjectures and Difference Fourier Transform,
Inventiones mathematicae, {\bf 125} (1996) 391, q-alg/9412016

\bibitem{MacConj} I.G. Macdonald, %Some conjectures for root systems,
SIAM J.Math. Anal. {\bf 13:6} (1982) 988-1007

\bibitem{CherednikDAHA} I. Cherednik, %Double Affine Hecke Algebras and Macdonald's Conjectures,
The Annals of Mathematics, Second Series, {\bf 141} (1995) 191-216

\bibitem{Koorn} T.H. Koornwinder, %Askey-Wilson polynomials for root systems of type BC,
in: {\sl Hypergeometric Functions on
Domains of Positivity, Jack Polynomials, and Applications (Tampa, FL, 1991)}, Contemp. Math. {\bf 138}, Amer. Math. Soc., Providence, RI, 1992, 189-204

\bibitem{Taki} M. Taki, JHEP {\bf 0803} (2008) 048, arXiv:0710.1776; arXiv:0805.0336

\bibitem{Che}  I. Cherednik, Internat.Math.Res.Notices, {\bf 1997 (10)} (1997) 449-467,
  % Difference Macdonald–Mehta conjecture, preprint,
  q-alg/9702022

\bibitem{EK}  P. Etingof, A. Kirillov, Electr.Res.Announc.Amer.Math.Soc. {\bf 4} (1998) 43-47,
q-alg/9712051
  % \textit{On Cherednik-Macdonald-Mehta identities} \\

\bibitem{ChaE}   O. Chalykh, P. Etingof, Advances in Mathematics, {\bf 238} (2013) 246-289,
  %% \textit{Orthogonality relations and Cherednik identities for multivariable Baker-Akhiezer
  %%   functions} \\
  arXiv:1111.0515

\bibitem{MMP} A.~Mironov, A.~Morozov, A.~Popolitov,
%``Cherednik-Mehta-Macdonald formula as a superintegrability property of a unitary model,''
Phys. Rev. \textbf{D110} (2024) 126026,
%doi:10.1103/PhysRevD.110.126026
arXiv:2410.03175

\bibitem{Wit}  E. Witten, %“Quantum Field Theory and the Jones Polynomial,” 
Commun. Math. Phys. {\bf 121} (1989) 351	


\end{thebibliography}
\end{document}